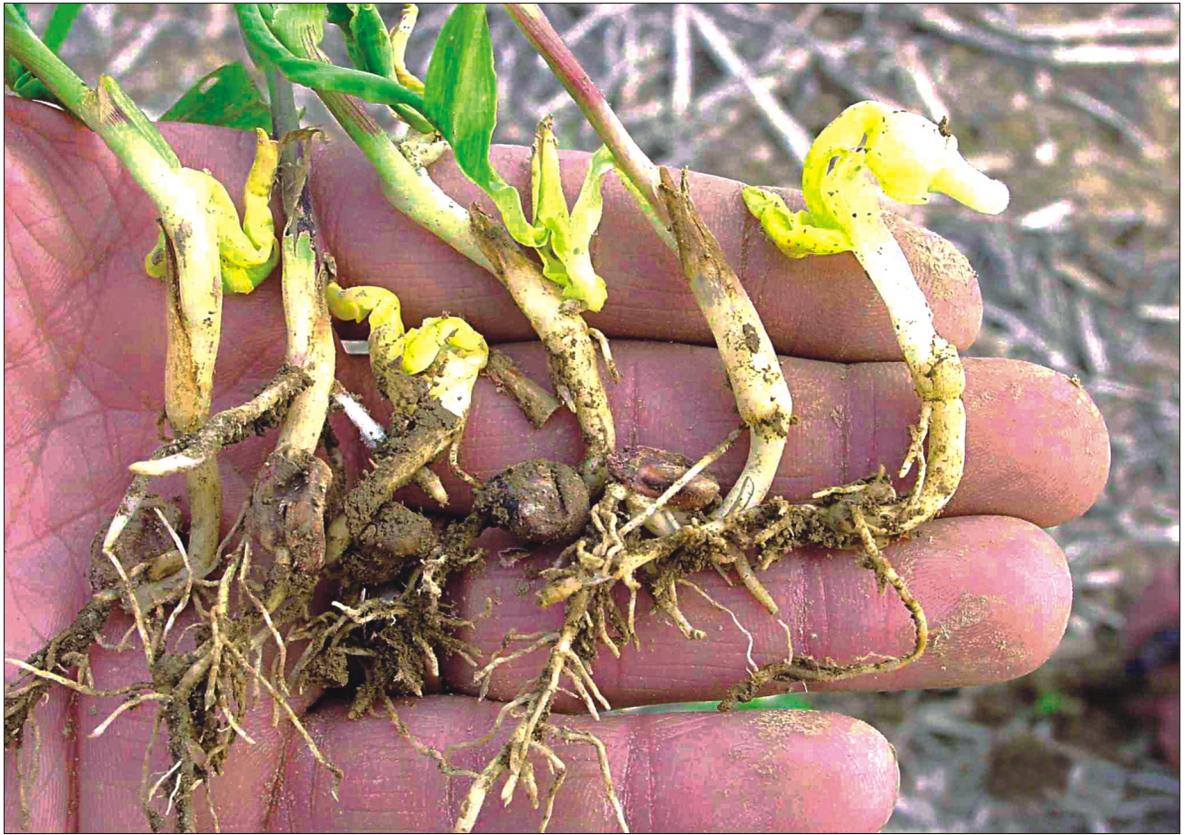

# Asian Journal of
# Plant Sciences

# Research Article
# Comparative Salt Tolerance Study of Some *Acacia* Species at Seed Germination Stage

[1]Chérifi Khalil, [2]Boufous El Houssein, [1]Boubaker Hassan and [1]Msanda Fouad

[1]Laboratory of Biotechnology and Valorization of Natural Resources, Faculty of sciences, Ibn Zohr University, P.O. Box 8106, 8000 Agadir, Morocco
[2]Department of Biochemistry and Microbiology, Laval University, Quebec City (Quebec), Canada

## Abstract

**Objective:** The purpose of this study was to assess and compare the seed germination response of six *Acacia* species under different NaCl concentrations in order to explore opportunities for selection and breeding salt tolerant genotypes. **Methodology:** Germination of seeds was evaluated under salt stresses using 5 treatment levels: 0, 100, 200, 300 and 400 mM of NaCl. Corrected germination rate (GC), germination rate index (GRI) and mean germination time (MGT) were recorded during 10 days. **Results:** The results indicated that germination was significantly reduced in all species with the increase in NaCl concentrations. However, significant interspecific variation for salt tolerance was observed. The greatest variability in tolerance was observed at moderate salt stress (200 mM of NaCl) and the decrease in germination appeared to be more accentuated in *A. cyanophylla* and *A. cyclops*. Although, *A. raddiana*, remains the most interesting, it preserved the highest percentage (GC = 80%) and velocity of germination in all species studied in this study, even in the high salt levels. This species exhibited a particular adaptability to salt environment, at least at this stage in the life cycle and could be recommended for plantation establishment in salt affected areas. On the other hand, when ungerminated seeds were transferred from NaCl treatments to distilled water, they recovered largely their germination without a lag period and with high speed. This indicated that the germination inhibition was related to a reversible osmotic stress that induced dormancy rather than specific ion toxicity. **Conclusion:** This ability to germinate after exposure to higher concentrations of NaCl suggests that studied species, especially the most tolerant could be able to germinate under the salt affected soils and could be utilized for the rehabilitation of damaged arid zones.

Competing Interest: The authors have declared that no competing interest exists.

Data Availability: All relevant data are within the paper and its supporting information files.

**INTRODUCTION**

Salinity of soils is one of the most environmental factors limiting agricultural production and has significant effects on crop productivity and biodiversity. It has more severe impact in arid and semi-arid environments and combined with the water constraint presents a serious threat to food stability in these areas[1-3]. Indeed, salinization already has affected more than 800 million hectares throughout the world or 6% of total land[4,5]. In the North Africa and the Middle East, salinity affects 15 million hectares of arable lands and this area is in continuous progression[6,7]. In Morocco, more than 5% of areas are already affected by salinity to various degrees[8]. They correspond especially to arid and semi-arid regions where 80% of available water for irrigation contains salinity[9] higher or equal to 5 g L$^{-1}$. These wide geographical area are not exploited to a great extent, except when they occasionally constitute a poor pasture land or irrigated domain with low yield.

The best economic approach for exploitation and rehabilitation of these marginal regions is selection of salt tolerant species and varieties capable of sustaining a reasonable yield within salt-affected soils[10-12]. The effectiveness of such approach depends on the availability of genetic variation in relation with salt tolerance and its exploitation by screening and selection of the powerful plants under saline stress[13-15]. For many crop species, intraspecific variability for salt tolerance have been identified among cultivars and wild species[16-19]. These rustic resources are well adapted and constitute a potential reservoir useful to provide interesting materials in order to diversify and increase the productivity, particularly in pasture land affected by salinity.

*Acacia* species have the ability to survive in a diverse range of habitat and environments. They are well adapted to the arid and semi-arid regions and are known for them tolerance to high pH and salinity as well[20,21]. These species are able to stabilize and fertilize soils via nitrogen fixing and mycorrhizal symbioses[22,23] and constitute sources of wood, fodder, medicine and gum[23]. *Acacia* trees may thus include ideal candidates for enabling saline land reclamation with a potential for financial returns because of their combined production and soil improvement[24-26].

In Morocco, the genus *Acacia* is represented by four spontaneous species including one Moroccan endemic (*Acacia gummifera* Wild.); the other three are *Acacia raddiana* Savi, *Acacia ehrenbergiana* Hayne., *Acacia albida* Del. Otherwise, about 10 species are introduced for ornamentation, reforestation or the fight against desertification[27].

Successful establishment of plants often depends on germination success, especially those that grow in salt affected areas. Thus, seeds must remain viable for long period in high salinity conditions and germinate when salinity decreases[28]. Various halophytic or glycophytic species, show an important variability with their ability to germinate under different salt concentration[29-31]. The effects on germination depend on the concentration of NaCl and varied among the plant species[32]. To overcome salt stress effect, plants have evolved various mechanisms that help them to adapt the osmotic and ionic stress caused by high salinity[25]. Otherwise, salts can affect germination of seeds either by restricting the supply of water (osmotic effect) or by causing specific injury through ions to the metabolic machinery (ionic effect)[33].

The purpose of this study was to assess and compare the seed germination response of 6 *Acacia* species (*Acacia gummifera*, *Acacia raddiana*, *Acacia eburnea*, *Acacia cyanophylla*, *Acacia cyclops* and *Acacia horrida*) under different NaCl concentrations in order to explore opportunities for selection and breeding salt tolerant genotypes that can be utilized in future land reclamation projects. The study will also assess to determine whether salt stress is induced by osmotic constraints or by toxic effect of NaCl.

**MATERIALS AND METHODS**

**Plant material:** Six *Acacia* species were studied: Tow spontaneous species (*Acacia gummifera* and *Acacia raddiana*) and four introduced species (*Acacia eburnea*, *Acacia cyanophylla*, *Acacia cyclops* and *Acacia horrida*). Mature, dry seeds were collected from trees growing under field conditions in semi-arid region of South Moroccan region. Seeds collected from pods were generously provided by the Regional Forest Seeds Station of Marrakech and stored in a cold chamber at 6°C.

*Acacia gummifera* is an endemic species to Morocco and *Acacia tortilis* (Forssk.) Hayne subsp., *raddiana* (Savi) Brenan, commonly named *A. raddiana*, considered as a keystone species is more prevalent in the inland part of the ecoregion and are widely distributed throughout the Sahara desert[34,35]. The other four species were introduced in Morocco from South-western Australian in the 18th century. These plantations were created for several purposes such as their use as ornamental plant, in the fight against desertification and for dune stabilization[36].

**Germination:** Seeds from different pods were manually scarified, to overcome hard seed coat dormancy and sterilized with 0.5% sodium hypochlorite solution (NaOCl) for 10 min, then rinsed with sterile distilled water several times and briefly blotted on filter paper. Three replicates of 20 seeds from each accession were placed in plastic petri dishes (90 mm diameter) on filter paper wetted with distilled water (control) and four salinity concentrations (100, 200, 300 and 400 mM NaCl). Petri dishes were randomized in a precision incubator and maintained in the dark at 25±0.5°C. Seeds were considered to have germinated when their radicle reached at least 3 mm long. Germination response was recorded daily for 10 days.

Several germination parameters were calculated to characterize the salt tolerance, including the corrected germination rate (GC), germination rate index (GRI) and mean germination time (MGT):

- Corrected germination rate (GC) was expressed as the number of seeds germinated in a concentration of salt divided by the number of germinated seeds in distilled water (control) for 10 days[37]
- The germination rate index (GRI) was calculated by using the following formula:

$$GRI = \left(\frac{G1}{1} + \frac{G2}{2}\right) + ... + \left(\frac{Gx}{x}\right)$$

where, G is the germination percentage at each day after sowing and 1, 2, ··· and x is the corresponding day of germination. The value of GRI was higher when seeds germinated earlier. This parameter described by Weng and Hsu[38] and Mirzamasoumzadeh *et al*.[39] is a measure of seedling vigor and should involve not only germination but emergence characteristics.

The Mean Germination Time (MGT) is a measure of the rate and time-spread of germination (lower values indicating faster germination). It was estimated as:

$$MGT = \sum \frac{ni \cdot ti}{N}$$

where, t is time from the beginning of the germination test in terms of days and n is the number of newly germinated seeds at time t[40,41].

To test germination recovery performance after salt exposure, ungerminated seeds in severe salt stress (300 and 400 mM of NaCl) were transferred to distilled water and incubated for 6 days. The recovery germination percentage was calculated by dividing the number of germinated seeds in recovery test by the number of seeds transferred to distilled water[42].

**Statistical analysis:** All values expressed as a percentage were arcsine square root transformed before performing statistical analysis to normalize the data and improve homogeneity of variance[43]. These measures were submitted to a two ways analysis of variance (ANOVA) with species and salinity treatments as factors followed by a Student-Newman-Keuls *post hoc* test. A difference was considered to be statistically significant when p<0.05. All statistical analysis were performed with Statistica software Version 6.1 for Windows[44].

## RESULTS AND DISCUSSION

**Effect of salinity on seed germination:** For the 6 species, the two-way ANOVA revealed highly significant main effect of both species and salinity regarding final germination percentage and germination rate index (p<0.001) (Table 1, 2). However the existence of a significant interaction between these two effects (F = 34.00 at (GC) and F = 6.385 at (GRI)) indicated that species studied did not similarly respond to the effect of salt at a given concentration of NaCl.

For each concentration of NaCl, a significant interspecific variation in both potential of seed germination (Fig. 1) and seedling vigor (Fig. 2) among the 6 examined species was observed. Mean comparison at different salinity levels indicated that increase of salinity causes a decrease in seed germination capacity, which was higher in distilled water than in any NaCl concentration. Moreover, all the species showed an increase in Mean Germination Time (MGT), indicating that seeds germinated more slowly as salinity increased (Table 3).

At the lowest stress (100 mM), the decrease in germination seemed to be more accentuated in *A. cyanophylla* (GC never fell above 30%). However, all the

Table 1: Two way analysis of variance (ANOVA) for final germination percentage (GC) of the 6 species under different concentrations of NaCl

| Source of variation | df | MC | F |
|---|---|---|---|
| Species (Sp.) | 5 | 0.618 | 270.110** |
| Salinity (Salt.) | 4 | 5.113 | 2233.421** |
| Sp.×Salt | 20 | 0.077 | 34.000** |
| Error | 60 | 0.002 | |

df: Degree of freedom, MC: Mean square, F: Ratio of variances, **Significant at 1% probability level

Table 2: Two way analysis of variance (ANOVA) for germination rate index (GRI) of the 6 species under different concentrations of NaCl

| Source of variation | df | MC | F |
|---|---|---|---|
| Species (Sp.) | 5 | 0.570 | 283.194** |
| Salinity (Salt.) | 4 | 4.149 | 2064.370** |
| Sp.×Salt | 20 | 0.040 | 20.010** |
| Error | 60 | 0.002 | |

df: Degree of freedom, MC: Mean square, F: Ratio of variances, **Significant at 1% probability level

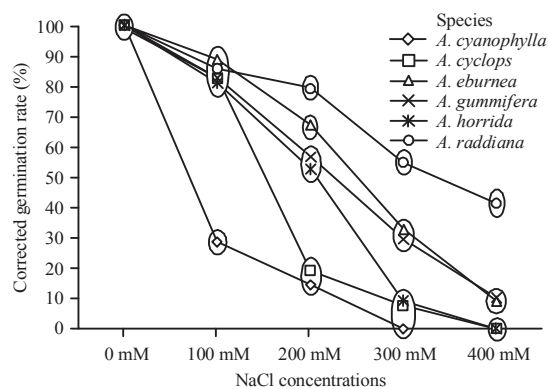

Fig. 1: Corrected germination rate of different species in various NaCl concentrations, at each concentration of NaCl, values in the same ellipse are not significantly different (p<0.05) (Newman-Keuls test)

Table 3: Mean Germination Time (MGT) of different species in various NaCl concentrations

| Species | NaCl concentrations (mM) | | | | |
|---|---|---|---|---|---|
| | 0 | 100 | 200 | 300 | 400 |
| A. cyanophylla | 2.48[a] | 2.98[a] | 5.83[b] | NC | NC |
| A. cyclops | 2.58[a] | 3.09[a] | 5.42[b] | 6.67[c] | NC |
| A. eburnea | 3.20[a] | 3.80[a] | 4.31[a] | 5.55[b] | 7.00[c] |
| A. gummifera | 3.05[a] | 3.48[a] | 4.24[a] | 5.66[c] | 6.22[c] |
| A. horrida | 3.03[a] | 3.58[ab] | 4.76[b] | 6.50[c] | NC |
| A. raddiana | 2.66[a] | 2.80[a] | 3.30[a] | 4.86[b] | 6.08[c] |

Values followed by the same letters in a row are not significantly different at p<0.05, (Newman-Keuls test), NC: MGT was not calculated because of insufficient germination other species are distinguished by forming a homogeneous group which is little affected by salinity compared with control (GC = 84% on average). At this saline level, seeds germinated rapidly and no significant change was noticed in germination speed (MGT not differing from that of control) (Table 3).

At 200 mM, the results revealed considerable interspecific variation in the response of seed germination to salinity among the studied species. Thus, four groups were distinguished, the first was formed by *A. raddiana* which showed very high germination percentage (GC = 80%) and revealed earlier seed germination MGT = 3.29, followed by *A. eburnea* (GC and GRI close respectively to 67 and 51%), while *A. gummifera* and *A. horrida* occupied an intermediate position between the tolerant species and the rest of the others, considered as the most sensitive (GC = 15% in the case of *A. cyanophylla* and 19% for *A. cyclops*). Otherwise, from this level, seed germination was relatively slower in *A. cyanophylla* and *A. cyclops* (high MGT).

The elevated doses of salt (300 and 400 mM) induced significant reduction in seed germination and retarded their initiation of all the species. The *A. cyanophylla*, *A. horrida* and *A. cyclops* appeared to be sharply affected in the same manner by these two concentrations, their corrected germination percentage does not exceed 6% and the time to germination gradually lengthened (MGT >6 days) at 300 mM. Moreover, no germination took place in this group at 400 mM. Seeds of *A. gummifera* and *A. eburnea* could be regarded as moderately tolerant to salt stress and reacted in the same way at these two concentrations (GC and GRI close respectively to 32 and 12% in the case of 300 mM). In this case germination was significantly delayed at the tow highest NaCl levels.

Whereas, *A. raddiana* continued to record highest percentage and velocity of germination in all species studied in this study, even in the high salt levels with GC reached 55% at 300 mM and 41% at 400 mM of NaCl. These species exhibited a particular adaptability to salt environment, at least at this stage in the life cycle and could be recommended for plantation establishment in salt affected areas.

The effect of salinity on germination has been addressed by several researchers and in different species[33,45-48]. This stage is very important for the development of the plants, particularly those that live in environments affected by salinity. In this study, the monitoring of germination process revealed that salinity was notably affected germination in *Acacia* species but also delayed the time needed to complete germination, especially with increasing salinity level. This is consistent with present result particularly for sensitive species and is explained by the time required for the seeds to develop mechanisms allowing to adjust them internal osmotic pressure[49].

Moreover, the results showed significant interspecific variation in salt tolerance during germination of the species studied in the range of concentrations of sodium chloride from 100-400 mM. This variability, required to start a breeding programs for salt tolerance, have been also observed in several species including halophyte or glycophyte species[15,42,48,50,51].

The Moroccan *Acacia raddiana* particularly, tolerated salinity until 400 mM (probably also at higher levels) with a germination rate that exceeded 40%. Indeed, according to Danthu *et al.*[52], *Acacia raddiana* is among the African *Acacia* species whose germination is less affected by the presence of salt. Its seed germination could be blocked only up to salt concentrations close to seawater (35 g L$^{-1}$). Previous studies have also reported that this species were the most tolerant and could be used in increasing forage production in salt affected areas[51,53-55]. Furthermore, the observed germination percentages in the current study were relatively well higher than those published by Jaouadi *et al.*[49] in Tunisian species (maximum germination rate of 21% under 22 g L$^{-1}$ of NaCl) and by Abari *et al.*[56] in Iranian taxa (germination was stopped at 300 mM). El Nour *et al.*[45] reported that no germination took place in four Yemeni *Acacia* (*A. cyanophylla*, *A. seyal*, *A. tortilis* and *A. tumida*) at salt concentration above 20 dS m$^{-1}$ (around 200 mM of NaCl).

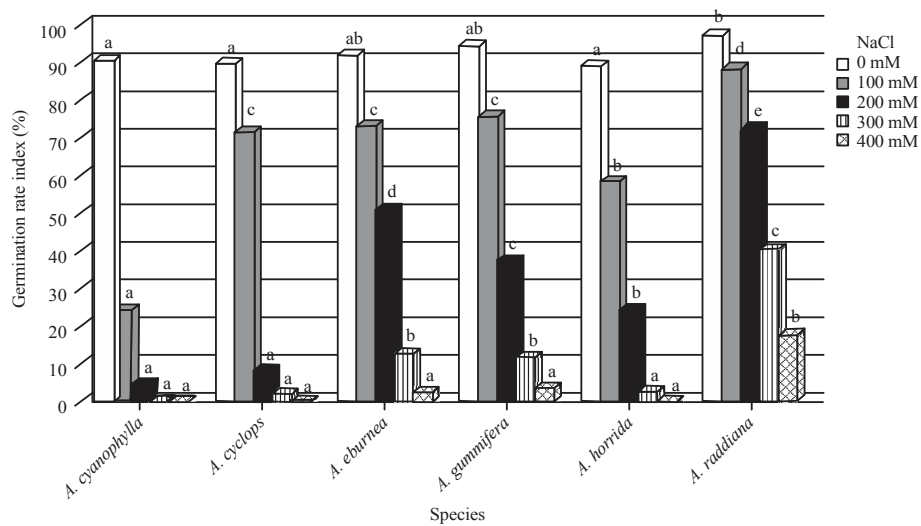

Fig. 2: Germination rate index of different species in various NaCl concentrations, at each concentration of NaCl, means of populations having the same letter are not significantly different (p<0.05) (Newman-Keuls test)

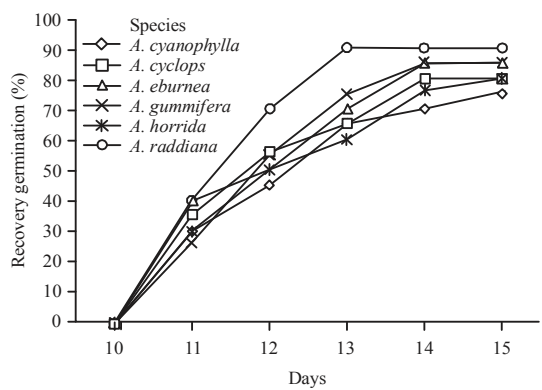

Fig. 3: Reversible effect of high NaCl concentration on the germination kinetics in the studied species

Among the rest of the 5 *Acacia* species examined in the present study, *A. gummifera* and *A. eburnea* was shown to possess the medium level of salt tolerance, whereas *A. cyanophylla*, *A. horrida* and *A. cyclops* were categorized as the least tolerant group. Several studies that have compared salt tolerance in some *Acacia* species have suggested relatively the same ranking[57].

Moreover, previous studies have reported that salt tolerance was positively correlated with seed size[58,25].

The *A. raddiana*, *A. gummifera* and *A. eburnea* seeds are larger than those of *A. cyanophylla*, *A. horrida* and *A. cyclops* and according to Croser *et al.*[59], larger seeds may contain more food reserves, which could be used to overcome osmotic effects of salts and greater energy reserves making them less dependent on photosynthesis for early growth. Similar observation which made on *Triticum aestivum*[60,61] on *Atriplex* species[62] as well as on *Acacia longifolia*[25], which also showed that larger seeds had greater success in overcoming osmotic constraints during the initial stages of germination.

**Recovery of germination:** After transfer of the seeds that failed to germinate under high concentrations of salt to distilled water, they recovered relatively their aptitude of germination at all the species studied (Fig. 3). The delay in germination speed tended to be relatively rapid than that observed in distilled water.

This, revealed that NaCl had no toxic effect, because salt stress did not damage the embryo as verified by not only the germination recovery but also pink embryos in the tetrazolium staining test. The NaCl concentrations did not destroy seed germination ability, it had only repressed the germination momentarily and the viability was preserved. These results are consistent with those obtained in several other salt tolerant species including *Atriplex halimus*[37], *Medicago ruthenica*[63], *Medicago polymorpha* and *Medicago ciliaris*[64], *Panicum turgidum*[29], *Triticum aestivum*[65], *Brassica napus*[19] and *Lolium perenne* and *Bromus tomentellus*[66].

An important characteristic of salt tolerant seeds, which differentiate them from seeds of glycophytes is their aptitude to maintain seed viability for lengthy period of time during exposure to hyper-saline conditions and then initiate germination when salinity stress is decreased[67,68,29]. The present study revealed that the recovery of germination is not a criterion of salt tolerance which distinguishes halophytes from glycophytes. It was maybe due to a reversible osmotic effect that induced dormancy, as revealed in findings of Khan and Ungar[69] and Tilaki *et al.*[66]. Consequently, a high proportion of seeds remained viable and had the ability to germinate when salinity stress was alleviated[29].

Reduction in mechanisms of germination by osmotic stress may be related to the lower diffusion of water through the seed coat[70,46] caused by the increased osmotic pressure environment, preventing the seed imbibition[62] and mobilization of reserves for embryo's growth[71]. Thus, dormancy decrease the risk of seedling mortality when moisture is limited and salinity is augmented[72]. High recovery germination speed observed in the studied species, particularly in *A. raddiana*, indicated that seeds have the ability to avoid deterioration caused by prolonged exposure to unfavorable biotic factors[73]. This situation constitutes an ecophysiological adaptive strategy to take full advantages of favorable conditions, available for a short time, during the germination stage[74]. It also, secure the long-term subsistence of seed bank helping the species in dispersal germination and seedling establishment over years[72]. Tilaki *et al.*[66] reported that under saline environments, seed survival may be a suitable condition for success instead of germination capacity, since recovery germination does occur in the seeds when hyper-saline conditions are alleviated.

## CONCLUSION

In the light of the above result, it is concluded that:

- Sodium chloride caused a reversible osmotic effect of germination rather than ion specific toxicity and exerts a temporary inhibition of germination which is eliminated with the removal of the constraint
- The ability to germinate after exposure to higher concentrations of NaCl suggests that Acacia species, especially the most tolerant, could be able to germinate under the salt affected soils and could be utilized for the rehabilitation of damaged arid zones
- The interspecific diversity in salt tolerance during germination is useful to start a breeding programs for salt tolerance
- The native *A. raddiana* species remained the most tolerant and conserve its aptitude to germinate until 400 mM, probably also at higher levels, consequently it was behaved as a halophyte plant
- More ambitious programs, including *Acacia* species are necessary, not only at germination but also at the other stages of the life cycle. This opens the possibility to continue this study to verify correlation between salt tolerance during seed germination and early stage of plant development which will be most useful in a breeding programs for selecting salt-tolerant in *Acacia* species

## ACKNOWLEDGEMENT

We express gratitude to the Regional Forest Seeds Station (R.F.S.S.M) of Marrakech (Morocco) for their information and guidance during prospecting and pods harvesting.